\documentclass[useAMS,usenatbib,psfig]{mn2e}
\usepackage{graphicx}
\usepackage{amsmath}
\usepackage{amssymb}
\usepackage{times}
\newcommand{\hii}{{H{\scriptsize II} }}
\newcommand{\aap}{A\&A}
\newcommand{\mnras}{MNRAS}
\newcommand{\apjs}{ApJS}
\newcommand{\apj}{ApJ}
\newcommand{\araa}{ARAA}
\title[Mid-Infrared Source Multiplicity within Hot Molecular Cores traced by Methanol Masers]{Mid-Infrared Source Multiplicity within Hot Molecular Cores traced by Methanol Masers}
\author[S.N.Longmore et al.]{S. N. Longmore$^{1,2}$, M. G. Burton$^{1}$, V. Minier$^{3}$ and A.J.Walsh$^{1}$\\
$^{1}$School of Physics, University of New South Wales, Sydney, 2052, NSW, Australia\\
$^{2}$Australia Telescope National Facility, CSIRO, Epping, 1710, Sydney, Australia\\
$^{3}$Service d'Astrophysique, DAPNIA/DSM/CEA Saclay, 91191 Gif-sur-Yvette, France\\
}
\begin{document}

\date{Accepted 2006 March 22.  Received 2006 February 6}

\pagerange{\pageref{firstpage}--\pageref{lastpage}} \pubyear{2006}

\maketitle

\begin{abstract}
We present high resolution, mid-infrared images toward three hot
molecular cores signposted by methanol maser emission; G173.49+2.42
(S231, S233IR), G188.95+0.89 (S252, AFGL-5180) and G192.60-0.05
(S255IR). Each of the cores was targeted with Michelle on Gemini North
using 5 filters from 7.9 to 18.5 $\mu$m. We find each contains both
large regions of extended emission and multiple, luminous point
sources which, from their extremely red colours ($F_{18.5}/F_{7.9}
\geq 3$), appear to be embedded young stellar objects. The closest
angular separations of the point sources in the three regions are
0.79, 1.00 and 3.33$\arcsec$ corresponding to linear separations of
1,700, 1,800 and 6,000AU respectively. The methanol maser emission is
found closest to the brightest MIR point source (within the assumed
1$\arcsec$ pointing accuracy). Mass and luminosity estimates for the
sources range from 3-22 M$_\odot$ and 50-40,000 L$_\odot$. Assuming
the MIR sources are embedded objects and the observed gas mass
provides the bulk of the reservoir from which the stars formed, it is
difficult to generate the observed distributions for the most massive
cluster members from the gas in the cores using a standard form of the
IMF. 
\end{abstract}


\section{Introduction}
Massive stars play a fundamental role in driving the energy flow and
material cycles that influence the physical and chemical evolution of
galaxies. Despite receiving much attention, their formation process
remains enigmatic. Observationally, the large distances to the nearest
examples and the clustered mode of formation make it difficult to
isolate individual protostars for study. It is still not certain, for
instance, whether massive stars form via accretion (similar to low
mass stars) or through mergers of intermediate mass stars.

Advances in instrumentation, have enabled (sub) arcsecond resolution
imaging at wavelengths less affected by the large column densities of
material that obscure the regions at shorter wavelengths. Recent
observations exploiting these capabilities have uncovered the
environment surrounding \emph{individual} massive protostellar
systems. From analysis of $\sim$2.3 $\mu$m CO bandhead emission,
\citet{BikThi2004} have inferred Keplerian disks very closely
surrounding (within a few AU) four massive young stellar objects,
while interferometric, mm-continuum observations, find the
mass-function of protostellar dust clumps lies close to a Salpeter
value down to clump radii of 2000AU \citep{BeutherSchilke2004}. These
high resolution observations point toward an accretion formation
scenario for massive stars.

Further discrimination between the two competing models is possible by
examining the properties, in particular the young stellar populations,
of hot molecular cores. The mid-infrared (MIR) window (7-25 $\mu$m)
offers a powerful view of these regions. The large column densities of
material process the stellar light to infrared wavelengths, and
diffraction limited observations are readily obtained.

Recent observations indicate that class \rm{II} methanol masers
exclusively trace regions of massive star formation~\citep{Minier2003}
and are generally either not associated or offset from UC\hii
regions~\citep{Walsh1998}. \citet{Minier2005} (hereafter M05) have
carried out multi-wavelength (mm to MIR) observations toward five star
forming complexes traced by methanol maser emission to determine their
large scale properties. They found that maser sites with weak
($<$10mJy) radio continuum flux are associated with massive
($>$50M$_{\odot}$), luminous ($>$10$^{4}$L$_{\odot}$) and deeply
embedded (A$_{v}>$40 mag) cores characterising protoclusters of young
massive (proto)stars in an earlier evolutionary stage than UC\hii
regions. The spatial resolution of the observations ($>$8$\arcsec$)
was, however, too low to resolve the sources inside the
clumps. Details of the regions from observations in the literature are
described in M05. We have since observed three of the M05 regions at
high spatial resolution to uncover the embedded sources inside the
cores at MIR wavelengths.

\section{Observations and Data Reduction}

The data were obtained with Michelle\footnote{The Michelle imager
employs a Raytheon 320x240 pixel Si:As IBC array with a pixel scale of
0.099$\arcsec$.} on the 8-m, Gemini North telescope in queue mode, on
the 18$^{th}$, 22$^{nd}$ and 30$^{th}$ of March 2003.  Each pointing
centre was imaged with four N band silicate filters (centred on 7.9,
8.8, 11.6 and 12.5 $\mu$m) and the Qa filter (centred on 18.5 $\mu$m)
with 300 seconds on-source integration time. G173.49 and G188.95 were
observed twice on separate nights and G192.60 observed once.

The N and Q band observations were scheduled separately due to the
more stringent weather requirements at Q band. The standard Chop-Nod
technique was used with a chop throw of 15$\arcsec$ and chop direction
selected from MSX images of the region, to minimise off-field
contamination. The spatial resolution calculated from standard star
observations was $\sim$ 0.36$\arcsec$ at 10 $\mu$m and $\sim$
0.57$\arcsec$ at 18.5 $\mu$m. The 32$\arcsec$x24$\arcsec$ field of
view fully covered the dust emission observed by M05 in each region.

Particular care was taken to determine the telescope pointing position
but absolute positions were determined by comparing the MIR data to
sensitive, high resolution, cm continuum, VLA images of the 3 regions
(Minier et al. in prep). Similar spatial distribution and morphology
of the multiple components allowed good registration between the
images. The astrometric uncertainty in the VLA images is
$\sim$1$\arcsec$.

Flux calibration was performed using standard stars within 0.3 airmass
of the science targets. There was no overall trend in the calibration
factor as a result of changes in airmass throughout the
observations. The standard deviation in the flux of standards
throughout the observations was found to be 7.4, 3.1, 4.4, 2.4 and 9\%
for the four N-band and 18.5 $\mu$m filters respectively. The
statistical error in the photometry was dominated by fluctuations in
the sky background. Upper flux limits were calculated from the
standard deviation of the sky background for each filter and a
3$\sigma$ upper detection limit is used in Table~1. Similarly, a
3$\sigma$ error value is quoted for the fluxes in Table~1 (typical
values for the N and Q band filters were 0.005 and 0.03 Jy
respectively). The flux densities for the standard stars were taken
from values derived on the Gemini South instrument,
T-ReCS\footnote{See http://www.gemini.edu}
which shares a common filter set with Michelle.

Regions confused with many bright sources were deconvolved using the
Lucy-Richardson algorithm with 20 iterations. This was necessary to
resolve source structure and extract individual source fluxes. The
instrumental PSF was obtained for each filter using a bright,
non-saturated standard star. The results were reliable and repeatable
near the brighter sources when using different stars for the PSF and
observations of the objects taken over different nights. As a further
check, the standard stars were used to deconvolve other standards and
reproduced point sources down to 1\% of the peak value after 20
iterations, so only sources greater than 3\% of the peak value were
included in the final images. The resulting deconvolutions are shown
in Fig 1.
\begin{center}
\begin{scriptsize}
   \begin{table*}
     \label{tab_sources}
	    \begin{tabular}{|c|c|c|c|c|c|c|c|c|c|c|c|}\hline\hline
	      
Source & R.A. & Dec &
\multicolumn{5}{c}{Flux (Jy)}&  T\tiny{col} & L\tiny{mir} & M & R \\

 &(J2000) & (J2000) & F$_{7.9}$ & F$_{8.8}$ &  F$_{11.6}$ & F$_{12.5}$ & F$_{18.5}$ & \small{K}& \small{ $10^{2} L_{\odot}$} & M$_{\odot}$ & \small{AU} \\\hline

G173.49:1 1$^{p,m}$ & 05 39 13.07 & 35 45 51.3 & 0.68$\pm$0.03 & 0.14$\pm$0.008  & 0.25$\pm$0.01 & 0.36$\pm$0.01 & 4.30$\pm$0.1& 102 & $2.8$ & 5 & 250 \\
G173.49:1 2$^{pe}$ & 05 39 13.02 & 35 45 50.7 & 0.32$\pm$0.02 & 0.04$\pm$0.004  & 0.06$\pm$0.005 & 0.07$\pm$0.007 & 1.90$\pm$0.08& 84 & $3.2$ & 5 & 390\\
G173.49:1 3$^e$ & 05 39 12.93 & 35 45 51.2 & 0.07$\pm$0.01 & 0.02$\pm$0.004  & 0.09$\pm$0.007 & 0.05$\pm$0.006 & 1.52$\pm$0.07& 81 & $2.9$ & 5 & 400\\
G173.49:1 4$^e$ & 05 39 13.25 & 35 45 52.6 & $<$0.006 & $<$0.002 & 0.01$\pm$0.004 & 0.05$\pm$0.007 & 2.27$\pm$0.09& 75 & $7.3$ & 7 & 740\\
\vspace{1mm}
(G173.49:1 5$^p$ & 05 39 13.34 & 35 45 49.6 & 0.04$\pm$0.007 & 0.02$\pm$0.004  & 0.02$\pm$0.003 & 0.02$\pm$0.004 & 0.03$\pm$0.01& 236 & $1.9$ & 5 & 3)\\
 
G188.95:1 1$^{p,m}$ & 06 08 53.34 & 21 38 29.0 & 2.62$\pm$0.06 & 0.55$\pm$0.02  & 2.19$\pm$0.03 & 3.83$\pm$0.05 & 18.44$\pm$0.3& 136 & $8.4$ & 7 & 240\\
G188.95:1 2$^p$ & 06 08 53.38 & 21 38 28.3 & 1.00$\pm$0.04 & 0.59$\pm$0.02  & 0.69$\pm$0.02 & 0.21$\pm$0.01 & 3.02$\pm$0.1& 97 & $3.6$ & 6 &  310\\
G188.95:1 3$^p$ & 06 08 53.42 & 21 38 29.2 & 0.02$\pm$0.01 & 0.01$\pm$0.005  & 0.23$\pm$0.01 & 0.71$\pm$0.02 & 1.84$\pm$0.08& 175 & $0.6$& 3 & 40\\
G188.95:1 4$^p$ & 06 08 53.49 & 21 38 30.5 & 0.32$\pm$0.02 & 0.03$\pm$0.004  & 0.21$\pm$0.009 & 0.49$\pm$0.02 & 1.34$\pm$0.06& 171 & $0.5$ & 3 & 40\\ 
G188.95:1 5$^p$ & 06 08 53.78 & 21 38 33.8 & 0.02$\pm$ 0.006& 0.01$\pm$0.003  & $<$0.004 & 0.01$\pm$0.002  & $<$0.03& $>$51 & -    & -   & -   \\
\vspace{1mm}
G188.95:1 6$^e$ & 06 08 54.15 & 21 38 25.6 & 2.09$\pm$0.06 & 1.02$\pm$0.03  & 2.80$\pm$0.04 & 2.37$\pm$0.04 & 23.64$\pm$0.4& 107 & $19$ & 9 & 590 \\

G192.60:2 1$^{pe,m}$ & 06 12 54.06 & 17 59 25.1 & 79.5$\pm$0.7 & 21.81$\pm$0.2  & 47.72$\pm$0.4 & 95.32$\pm$1.0 & 76.79$\pm$2.4& 388 & $75$ & 13  & 90 \\
G192.60:2 2$^p$ & 06 12 53.90 & 17 59 25.7 & 6.62$\pm$0.1 & 3.9$\pm$0.05  & 7.88$\pm$0.08 & 11.09$\pm$0.1 & 16.27$\pm$0.5 & 240 & $72$ & 13  & 80 \\
G192.60:2 3$^e$ & 06 12 54.95 & 17 59 21.3 & 0.81$\pm$0.03 & 1.04$\pm$0.02  & 0.86$\pm$0.02 & 0.40$\pm$0.02 & 31.93$\pm$1.0& 67 & $430$ & 22 & 7060 \\
G192.60:2 4$^e$ & 06 12 54.73 & 17 59 32.3 & $<$0.005 & $<$0.002 & $<$0.003 & $<$0.002 & 3.2$\pm$0.13& $<$44 & -  & -    & -  \\
G192.60:2 5$^e$ & 06 12 54.39 & 17 59 26.8 & $<$0.005 & $<$0.002 & $<$0.003 & $<$0.002 & 0.9$\pm$0.06& $<$51 & -  & -    & -  \\

 \hline

	  \end{tabular}
	    \caption{Positions, measured fluxes and derived properties
	          of the detected mid-infrared objects in the three
	          cores.  T$_{col}$ is the colour temperature of the
	          source derived from the 12.5 and 18.5 $\mu$m flux
	          ratios. R and $L_{mir}$ are the radius and
	          luminosity of a greybody at temperature $T_{col}$.
	          The table uses notation Gxxx.xx:a~b for the sources,
	          where Gxxx.xx is the formation region from M05, `a'
	          is the clump number from M05 and `b' is the
	          mid-infrared source number.  The superscript p, pe
	          and e in the first column, denote point sources,
	          point sources with low level surrounding extended
	          emission and extended sources respectively. The
	          superscript m denotes the source closest in
	          projection to the methanol maser. The radius in the
	          final column is from the black body fit and is
	          consistent with observations.  The quoted errors and
	          upper flux limits are 3$\sigma$ values. Distances of
	          1.8, 2.2 and 2.6 kpc are used for G173.49, G188.95 and
	          G192.60 respectively. }
	    
     \end{table*}
\end{scriptsize}
\end{center}

\vspace{-1.2cm}
\section{Deriving Physical Parameters}
A colour temperature was derived for each object, assuming the flux
was greybody emission from dust at a single temperature and emissivity
given by 1 $-$ $e^{-\tau_\lambda}$. $\tau_\lambda$ was calculated
using A$_v$ (from M05) and the~\citet{Draine1989} normalised silicate
profile ($\tau_{12\mu m}$=$\tau_{18\mu m}$=$0.022Av$). The 12.5 to
18.5 $\mu$m flux ratios were used to derive the temperature as they
are expected to contain the least PAH emission and are less affected
by the broad-band silicate absorption features. In cases where either
the 12.5 and 18.5 $\mu$m source flux was below the detection limit,
only an upper and lower temperature limit respectively was
derived. Luminosities were derived in the blackbody limit at the
calculated source colour temperature. The values agreed to $<$1\% with
those calculated by integrating the product of the Planck function and
the emissivity function from 1-600$\mu$m so provide a reasonable lower
limit to the bolometric luminosity~\citep{DeBuizer2005}. Luminosities
derived from colour temperatures agree well with those from radiative
transfer model fits to the SED across the near and MIR (De Buizer,
priv. comm.). A mass estimate for the sources was calculated by
considering the range of plausible $\alpha$ values in a mass
luminosity-relationship, L$\propto$M$^\alpha$, for a deeply-embedded,
zero age main sequence star. For stars on the main sequence
$<$20M$_\odot$, $\alpha$=3.45 (eg~\citet{Allen1975}), while for stars
$>$ 20M$_\odot$, $\alpha \sim$3~\citep{Zinnecker2003}. Additional
luminosities due to accretion (accounting for up to half the total
luminosity~\citep{Osorio1999}) may increase $\alpha$ up to
$\sim$4. Over this range of $\alpha$ (3$<$$\alpha$$<$4), a factor of
ten error in luminosity corresponds to only a factor of $\sim$2 in the
calculated mass. An intermediate value, $\alpha$=3.45, was chosen to
derive the mass estimate. Flux errors of 20\% and a 20\% distance
uncertainty corresponds to a colour temperature error of 10\% and
luminosity error of 35\%. However, the relative luminosity uncertainty
for sources within a core drops to 20\% as they lie at the same
distance. Using the 3$\sigma$ upper flux limits from the 12.5 and
18.5$\mu$m images the lower mass detection limit is $\sim$
1M$_{\odot}$ for an object at a typical observed source temperature
of 130K at 1.8 kpc. The source positions, fluxes and derived properties are shown
in Table~1.

\vspace{-3mm}

\section{Results and Discussion}
Figure 1 shows the 18.5 and 7.9 $\mu$m images of the three
regions. The insets show deconvolved 7.9 $\mu$m images toward bright,
confused regions. All the sources toward each region are assumed to
belong to the same protocluster due to their sharply rising MIR flux
($F_{18.5}/F_{7.9} \gtrsim 3$) and large optical depths ($A_{v}>40$
mag). Although $F_{18.5}\sim F_{7.9}$ for G192.60:2 1, the spectral
energy distribution peaks in the MIR so it must still be heavily
embedded. Due to the low signal to noise of the detection G173.49:1 5
was not included in the following analysis.

Table~1 lists the positions, morphology, measured fluxes and derived
properties of the detected MIR sources and Table~2 shows the large
scale clump properties derived in M05. In all three clumps, the
individual source radii derived from the black body assumption compare
well to the observed radii. The variation in individual source radii
and temperature compared to the M05 values in Table~2 is not
unexpected due to the M05 assumption of a single powering source.
With a factor of $\sim$2 error in the M05 luminosity (see
eg~\citet{Purcell2006} for error analysis of two-component, greybody
fitting) the total measured MIR luminosity (L$_{mir}$) for G188.95 and
G192.60 accounts for the M05 total core luminosity (L$_{tot}$). The
large discrepancy in G173.49, however, suggests there are undetected,
luminous, cold source(s). Attributing the missing flux to a single
source would correspond to a star of mass $\sim$20M$_\odot$.\\

\vspace{-3mm}

\subsection{Nature of the MIR sources} 
MIR emission traces dust heated by nearby sources. Its morphology
provides clues as to the nature of the source, in particular, whether
the dust is heated externally or by an embedded source.
Observationally, the sources fall into three morphological categories;
(i) unresolved point sources, (ii) unresolved point sources with weak
surrounding extended emission and (iii) extended emission, which we
have designated in Table~1 as P, PE and E, respectively. The source
morphology is consistent across the filter range (after deconvolution
at longer wavelengths) but it is unclear whether G173.49:1 3 is
actually a colder component of the extended emission surrounding
G173.49:1 2 instead of a separate source.

 With a spatial resolution $\sim$400AU we have made the assumption
that the point-like sources (P and PE) are stellar in origin and the
extended sources (E) are non-stellar (eg externally heated) with the
exception that the most luminous source in the region is expected to
be internally heated regardless of morphology. The internal heat
sources are expected to be either very young stars or protostars. All
non-stellar sources have been excluded from the analysis in $\S\S$4.2,
4.3 and 4.4.

  \begin{table}
    \label{m05_clump_properties}
     \begin{center}
       \begin{tabular}{|c|c|c|c|c|c|c|c|c|}\hline\hline
	 
	 Clump & $L_{tot}$      & $L_{hot}$    &  $L_{mir}$     & $M_{gas}$       &  $T_{hot}$   & $R_{hot}$    & N{\scriptsize H$_2$}\\ 
	      & $10^{4}$       & $10^{2}$      &  $10^{2}$      &  $10^{2}$       & (\small{K})  & (\small{AU}) & 10$^{22}$           \\
	      & ($L_{\odot}$)  & ($L_{\odot}$) &  ($L_{\odot}$) &  (M$_{\odot}$)  &              &              & (cm$^{-2}$)         \\
	 \hline
	 G173.49:1 & 4.9 & 3.7 & 18  & 1.2 & 122 & 200   & 16\\ 
	 G188.95:1 & 1.1 & 77  & 32  & 0.5 & 150 & 600   & 5\\ 
	 G192.60:2 & 5.1 & 390 & 577   & 3.2 & 225 & 600 & 22\\
	 \hline
	 
       \end{tabular}
       \caption{Properties of the three cores derived from M05 using a
	        greybody approximation to fit the spectral energy
	        distribution.  L$_{tot}$ is the total clump
	        luminosity, L$_{hot}$, T$_{hot}$ and R$_{hot}$ give
	        the luminosity, temperature and radius of the hot
	        component derived from the greybody
	        approximation. L$_{mir}$ is the total luminosity of
	        the sources in each core from Table~1. M$_{gas}$ is
	        the average gas mass derived from mm-continuum and
	        molecular spectral line observations of the
	        clumps. The N{\scriptsize H$_2$} column densities are
	        the values derived from mm-continuum observations in
	        M05.  }
      
     \end{center}
     \vspace{-2mm}
  \end{table}

\subsection{Point source spatial distribution}  
As the large scale clump dust and gas morphology appears simple and
centrally peaked (see M05), we make the reasonable assumption that the
protocluster centres coincide with the central peak of dust
emission. The spatial distribution of the point sources within the
protocluster is similar between the clumps with close point sources
toward the cluster centre. The methanol masers are found closest to
the brightest MIR point source (within the assumed 1$\arcsec$ pointing
error from image registration). These sources have temperatures
sufficient to evaporate methanol ice from the dust grains into the gas
phase ($>$90K) as well as sufficient luminosity of IR photons to pump
the masing transition -- conditions models suggest are required for
such emission~\citep{Voronkov2005}.

It is known that more massive stars favour cluster centres
(e.g.~\citet{deGrijs2002}), but it is unclear whether they form there
or migrate in from outside. We have used the simple-harmonic model of
ballistic motion developed by~\citet{Walsh2004} to consider the
motion of sources within the cores. Using the measured column density
and radius from M05 (listed in Table~2), the time required for
migration from the edge to the centre is $\sim$ $ 10^5$ years. This is
comparable to the predicted HMC lifetime of 10$^5$ years
\citep{Kurtz2000} so we can not rule out the possibility of migration
within the clumps. Any sources having migrated to the centre in this
way would have acquired a velocity of $\sim$~2~kms$^{-1}$ with respect
to the clump.

\subsection{MIR source multiplicity}
Massive stars in clusters are observed to have a high companion star
fraction~\citep{Zinnecker2003}. In the M16 cluster,
\citet{Duchene2001} observed massive stars (earlier than B3) with
visual companions separated by 1000-3000AU. If multiple systems are
bound from birth, it is likely some of the sources we have observed
will belong to multiple systems, even though the companions may lie
below the detection limit. However, all three regions show two or more
point sources at close angular separation (see insets of Figure 1)
corresponding to linear separations of 1700 to 6000AU.  We cannot
determine whether these stars are physically bound or simply close due
to projection effects but we can calculate the instrumental
sensitivity required to confirm or deny the association. Assuming they
are physically bound in a Keplerian orbit, the maximum proper motions
(projection angle = 0$^\circ$) of $\sim$ 0.1 mas/year are too small to
be detected on short temporal baselines.  The maximum velocity
difference (projection angle = 90$^\circ$) $\sim$ 2 kms$^{-1}$ is
achievable by high spectral resolution observations of any line
features.

\subsection{Protocluster mass distribution}
The mass distribution of stars is generally well described as a power
law through the initial mass function (IMF).  Given the mass of gas
available to form stars, we may estimate the likelihood that a cluster
will end up with the most massive stars that are observed in it. The
fraction of gas that forms stars is given by the star formation
efficiency (SFE) and is observationally found to be less than 50\% in
any cloud and to be $\leq$ 33\% for nearby embedded clusters
\citep{2003LadaLada}. For a cluster whose total stellar mass is 120
(50, 320) M$_{\odot}$ (equivalent to the gas mass determined for the
three cores), \citet{2004WeidnerKroupa} estimate that the mean maximum
mass that a star may have in it is 10 (5, 20) M$_{\odot}$. This is
comparable to the largest observed mass in two out of the three
cases. However, we also observed several other stars in each cluster
so can estimate the probability of generating stars of equal or
greater mass than the remaining mass distribution. We did this by
running Monte-Carlo simulations to populate 10$^5$ clusters using
\citet{Salpeter1955}, \citet{MillerScalo1979} and \citet{Scalo1986}
IMFs until the available gas mass was exhausted. We only considered
clusters which contained a star of at least equal mass to the most
massive observed. The simulations show that even using the Salpeter
form of the IMF (most biased toward forming high-mass stars) and
allowing 50\% of the gas to form stars, it is difficult to generate
the observed mass distributions (probabilities $\lesssim$ 10$^{-2}$,
10$^{-5}$, 10$^{-1}$ for the three cores respectively). By itself,
this may not be significant for a single cluster. However, since the
probability is low for all three sources studied, it is unlikely that
the mass distribution of the most massive stars can be produced by
sampling a standard form of the IMF from the reservoir of gas
available for star formation. This conclusion would not hold if there
was a substantial stellar mass already in the cluster that remains
unseen, or if much of the original gas mass had already been dispersed
from the core due to star formation. The former requires a SFE close
to unity and given the relatively quiescent state of the cores, the
latter seems unlikely.

\vspace{-4.6mm}

\section{Conclusions}
A larger sample of young, massive protoclusters is required to draw
general conclusions. However, in all three hot molecular cores traced
by methanol maser emission we have found:
\begin{itemize}
\item Multiple, MIR sources which can be separated into three
  morphological types: unresolved point sources (P); unresolved point
  source with weak surrounding extended emission (PE) and extended
  sources (E). 
\item The point sources lie at close angular separations. Future high
  spatial and spectral resolution observations may be able to
  determine whether or not they are physically bound.
\item The methanol masers are found closest to the brightest MIR point
source (within the assumed 1$\arcsec$ pointing accuracy).
\item Cooler, extended sources dominate the luminosity.
\item The time scale for a source at the core edge
to migrate to the centre is comparable to the Hot Molecular
Core lifetime, so it is not possible to rule out large protostellar
motions within the core.
\item From the derived gas mass of the core and mass estimates for the
sources, Monte Carlo simulations show that it is difficult to generate
the observed distributions for the most massive cluster members from
the gas in the core using a standard form of the IMF. This conclusion
would not hold, however, if most of the original gas has already
formed stars, or has been dispersed such that the original core mass
is much greater than now observed.
\end{itemize}

\section{Acknowledgments}

S.L. would like to thank Alistair Glass, Scott Fisher, Tony Wong and
Melvin Hoare for helpful discussion of the data and scientific
input. We thank the anonymous referee for the thorough response and
insightful comments. This work was made possible by funding from the
Australian Research Council and UNSW. The Gemini Observatory is
operated by the Association of Universities for Research in Astronomy,
Inc., under a cooperative agreement with the NSF on behalf of the
Gemini partnership: NSF (USA), PPARC (UK), NRC (Canada), CONICYT
(Chile), ARC (Australia), CNPq (Brazil) and CONICET (Argentina).

\begin{figure*}
\label{fig:image_and_sed}
\begin{center}
  \includegraphics[width=0.95\textwidth, angle=0, trim=0 0 -5 0]{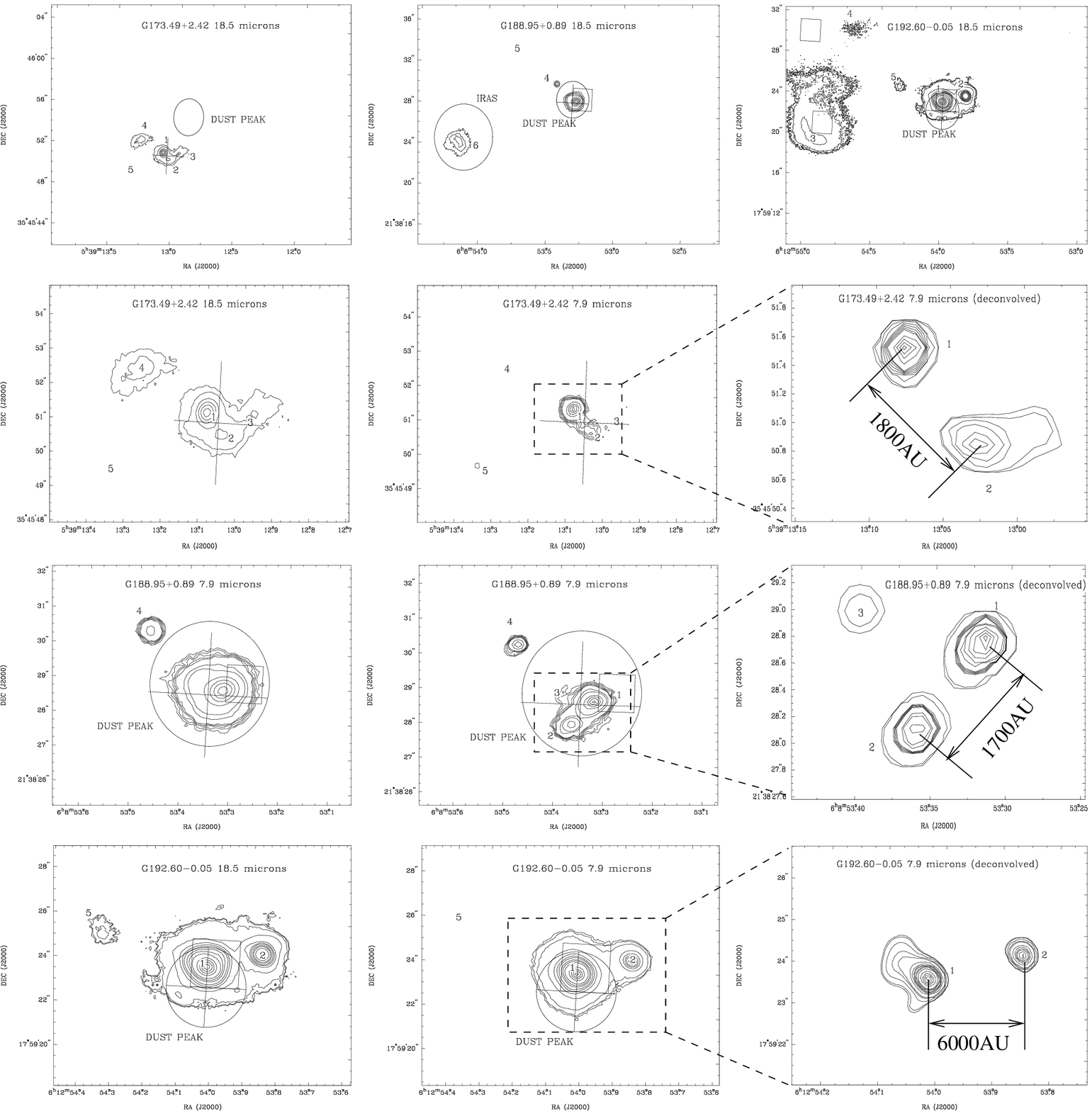}
  \caption{Images of the three regions. The top row displays the full
  field of view at 18.5 $\mu$m toward G173.49+2.42 Clump 1 (left),
  G188.95+0.89 Clump 1 (centre) and G192.60-0.05 Clump 2 (right). The
  centre of the mm dust peak emission from M05 is shown as an ellipse,
  methanol maser sites are shown as plus symbols and radio continuum
  sources as boxes. A further ellipse in the G188.95+0.89 images shows
  the position of the IRAS point source within the field but
  registration with a single MIR source is not possible due to the
  much larger IRAS beam size. The next three rows show close up images
  at 18.5 $\mu$m (left) and 7.9 $\mu$m (centre) of the bright point
  sources close to the dust peak and methanol maser emission
  (G173.49+2.42 (second row), G188.95+0.89 (third row) and
  G192.60-0.05 (bottom row)).  The inset on the right shows the result
  when the boxed region in the 7.9 $\mu$m image is deconvolved with
  the instrument PSF. Scale bars in the deconvolved images give the
  linear separation of the sources assuming they lie at the same
  distance.  }
\end{center}
\end{figure*}
\end{document}